\documentclass[manuscript]{copernicus}  

\nolinenumbers
\usepackage{color}
\usepackage[T1]{fontenc}

\begin{document}

\title{{Mirror mode junctions as sources of radiation }  
}

\author[1]{R. A. Treumann$^*$}
\author[2]{Wolfgang Baumjohann}

\affil[1]{International Space Science Institute, Bern, Switzerland}
\affil[2]{Space Research Institute, Austrian Academy of Sciences, Graz, Austria 

$^*$On leave from Department of Geophysics and Environmental Sciences, Munich University, Munich, Germany

\emph{Correspondence to}: W. Baumjohann (Wolfgang.Baumjohann@oeaw.ac.at)}

\runningtitle{Mirror mode junction}

\runningauthor{R. A. Treumann, W. Baumjohann}

\received{ }
\revised{ }
\accepted{ }
\published{ }


\firstpage{1}

\maketitle

\noindent\textbf{Abstract}. -- 
Mirror modes in collisionless high-temperature plasmas represent macroscopic high-temperature quasi-super\-con\-duc\-tors with bouncing electrons in discrete-particle resonance with thermal ion-sound noise contributing to the ion-mode growth beyond quasilinear stability. In the semi-classical GL approximation the conditions for phase transition are given. The quasi-superconducting state is of second kind causing a magnetically perforated plasma texture. Focussing on the interaction of mirror bubbles we apply semi-classical Josephson conditions and show that a mirror perforated plasma emits weak electromagnetic radiation which in the magnetosheath is in the sub-millimeter respectively infrared range.


\vspace{0.5cm}
\section{Introduction}
The mirror mode  \citep{chandrasekhar1961,vedenov1961,hasegawa1969,hasegawa1975,davidson1972,kivelson1996,gary1993,southwood1993,pokhotelov2001,pokhotelov2000,pokhotelov2002,pokhotelov2004,constantinescu2002,constantinescu2003,sulem2011,rincon2015,noreen2017} which, in high temperature plasma, evolves under anisotropic $A_i=P_{i\perp}/P_{i\|}-1>0$ pressure conditions, can be interpreted as kind of a phase transition from (unstable) normal to (stationary) second-kind quasi-superconducting state \citep{treumann2019,treumann2020}. It causes the plasma to become magnetically perforated. This raises the question, investigated in this letter, in what way closely spaced mirror bubbles may interact possibly producing identifiable effects other than diamagnetic field depletions. 

This second-kind superconducting phase transition \citep{ginzburg1950} is known from low temperature solid state physics \citep{bardeen1957,callaway1990}, evolving Meissner diamagnetism based on electron pairing and condensation that pushes the magnetic field locally out. In mirror modes the possibility of similar condensations has recently been demonstrated \citep{treumann2019}. The transition is initiated by the mirror instability, which starts under the necessary condition of positive ion pressure anisotropy $A_i\neq0$. In addition, the sufficient  condition for instability requires  the magnetic field strength $B$ to drop below a  threshold $B_c$ 
\begin{equation}\label{eq-bc}
B<B_c\approx \sqrt{2\mu_0NT_{i\perp}A_i}\,|\sin\theta_m|
\end{equation}
which follows from the linear ion-mirror growth rate
\begin{equation}\label{eq-gamma}
\frac{\gamma}{\omega_{ci}}\approx \frac{k_\|\lambda_i}{1+A_i}\sqrt{\frac{\beta_{p\|}}{\pi}}\Big[A_i+\sqrt{\frac{T_{e\perp}}{T_{i\perp}}}A_e-\frac{k^2}{k^2_\perp\beta_{p\perp}}\Big]
\end{equation}
The electron anisotropy $A_e\approx0$ is assumed negligible initially, $k_\|\ll k_\perp, \lambda_i=c/\omega_i$ ion inertial length, plasma-$\beta_p=2\mu_0N(T_i+T_e)/B_0^2$. Sometimes  this is inverted into a condition on the (arbitrary) angle of wave propagation \citep{southwood1993} $\theta_m=\tan^{-1}{k_\perp/k_\|}$ but the relevant physics is contained in the field threshold. The two conditions together yield that the critical (ion) temperature is $T_{ci\perp}=T_{i\|}$ which means  that 
\begin{eqnarray}\label{eq-2}
T_{i\|}-T_{i\perp} &\geq& 0\qquad \mathrm{stability, ``normal"~state} \nonumber\\[-2ex]
&&\\[-2ex]
T_{i\|}-T_{i\perp} &<& 0\qquad \mathrm{instability, ``quasi-superconducting"}\nonumber
\end{eqnarray}
which indeed reminds at the solid state superconducting phase transition. The mirror instability  readily saturates quasilinearly on the expense of the ion anisotropy \cite[cf., e.g.,][]{davidson1972,treumann1997} forming elongated  $k_\|\ll k_\perp$ magnetic bottles. Landau diamagnetic theory  \citep[cf., e.g.,][]{huang1973} suggests that any finite temperature diamagnetism is macroscopically very small, which is confirmed by simulations \citep{noreen2017} which show the saturation amplitude to remain minuscule. The observation of  large-amplitude localized quasi-stationary magnetic depletions of $\lesssim50\%$ \citep[see][for examples in high resolution]{treumann2018} must be enforced by conditions which are not included in linear or quasilinear theory \citep{treumann2004}. We do not go into discussing this problem here as it has been the subject of previous publications \citep[cf.,][]{treumann2018,treumann2019}. We just note that a number of simulations \citep{rincon2015,yao2019} and theory \citep{sulem2011} claim that nonlinear interactions between ions and waves provide large magnetic amplitudes \citep[see the discussion in][]{treumann2019}. In simple words, nonlinear scattering off waves, as sometimes assumed, increases diffusivity and entropy which primarily inhibits structure formation. The argument of increased internal pressure also fails because pressure is compensated by the large and elastic environment volume. It  dilutes the magnetic field only infinitesimally. Bubbles can deepen only on the expense of their neighbours along the same flux tube by sucking in plasma, which contradicts the observation of long mirror chains while supporting observation of isolated bubbles \citep{luehr1987,treumann1990}.

\section{Quasi-superconducting phase transition}
One way out of the above mentioned basic physical dilemma between observation and theory may be related to the resonance of bouncing particles  in the mirror bubble and the persistent thermal ion-acoustic background noise which is independent of the presence of mirror modes \citep{treumann2018,treumann2019}. These resonant bouncing particles (we here restrict to electrons, but ions if bouncing could contribute in a similar way as well) form the required condensate for phase transition. 
\subsection{Condensate formation}
The resonance is a discrete particle effect. It applies to all electrons in the Debye sphere and is in contrast to the small number of Landau-resonant electrons \citep[like in the radiation belts,][]{kennel1966} which generate the banded whistler lion roars \citep[cf., e.g.,][]{smith1976,tsurutani1982,zhang1998,baumjohann1999,maksimovic2001,ahmadi2018,breuillard2018,giagkiozis2018} in mirror bubbles. In contrast, trapped electrons performing their bounce motion in the quasilinearly stable mirror bubble resonate with the permanently present thermal ion-acoustic background noise of frequency $\omega_s$ through $v_\|\approx \omega_s/k$ near their mirror points where their parallel velocity $v_\|\approx c_s$ becomes comparable to the ion-sound speed $c_s=\sqrt{T_e/m_i}$. This applies to a large number $N_p$ of electrons. In resonance they become locked to the wave and drop out of their bounce motion while maintaining their large energy anisotropy. Many such locked electrons form the condensate. Their anisotropy further increases when they move with the ion-acoustic wave into the strong magnetic field beyond their mirror points. This mechanism also causes a weak attractive potential field at distance $\xi\approx 1.5\lambda_D$ in the wake of each resonant electron which by trapping another electron acts as Pippard's correlation length. The fractional number density of condensate electrons $\mathcal{N}=N_p/N_0<1$ may in this case not be small. Under the assumption that initially $A_e\approx 0$, the condensate anisotropy becomes 
\begin{equation}
A_e=\frac{2T_{e}}{m_ec_s^2}-1=\frac{2m_i}{m_e}-1\approx \frac{2m_i}{m_e}
\end{equation}
which enters the above ion-mode growth rate at quasilinear stability with quasilinearly compensated ion contribution, causing the instability to grow beyond the quasilinear limit. This effect corresponds to the noted condensate formation by pairing in metals though is basically different as here it is a high temperature classical effect. Further evolution implies pressure balance and the cause of surface gradient currents which has the effect of generating a partial London-Meissner diamagnetic effect. This phase transition can be treated in analogy to Ginzburg -Landau theory \citep{treumann2020}. It is, here, instructive to note that the above noted similar discrete-particle ion-resonance with ion-sound waves, though possible, as can be easily shown in the same way as for electrons, yields that the condensate-ion anisotropy becomes $A_i\approx T_i/T_e-1$. This ion-condensate anisotropy is positive only in high ion temperature plasma $T_i>T_e$ where ions have become heated, for instance by the presence of a shock and behind it, as would be the case in the magnetosheath. However, in order to compete with the electron anisotropy $T_i\gg T_e$ is required which is probably unrealistic. 

\subsection{GL-theory}
The semi-classical first Ginzburg-Landau equation is obtained by putting $\hbar\to0$. With $\mathbf{A}$ magnetic vector potential it reads
\begin{equation}
\frac{e^2}{2m_e}\mathbf{A}^2\psi+\alpha\psi+\beta|\psi|^2\psi=0
\end{equation}
where $\psi$ is the condensate wave function, $\alpha,\beta$ constants. Clearly $\beta>0$. It has the symmetric normal state solution $\psi=0$ for $\alpha>0$, and in condensate formation $\alpha=-a<0$ after symmetry breaking the solution
\begin{equation}
\beta|\psi|^2=a-\frac{e^2}{2m_e}\mathbf{A}^2>0
\end{equation}
which means that the density $N_p$ of resonant electrons is finite. Here, $|\psi|^2\sim N_p/N_0\equiv\mathcal{N}$ as we will also use it below when discussing radiation. Hence phase transition occurs if only $a>(e^2/2m_e)\mathbf{A}^2$. In the magnetosheath we have $|\mathbf{A}|\sim B\Delta L\approx 10^{-3}$ Vs/m. Hence the requirement is that 
$a\gtrsim 10^{-14}$ VAs or $\sim10^5$ eV. (One may note that for ion condensate formation this value reduces by more than three orders of magnitude and might thus favour ions even though their anisotropy cannot compete with that of electrons, a case which we do not investigate further here.)
Since we require that $\mathcal{N}<1$ the condition on the coefficients in GL-theory is simply that 
\begin{equation}
1<a/\beta<1+(\beta/10^5~\mathrm{eV})^{-1}
\end{equation}
The absolute values of these coefficients are unimportant. Approximate relations between these coefficients and the mirror conditions have also been obtained \citep{treumann2020} but will not be repeated here. Hence there is some range where phase transition becomes probable which, for the purpose of this Letter, should suffice. The physical meaning is that the discrete resonant-electron condensate causes macroscopic diamagnetism which substantially diminishes the magnetic field locally.

\subsection{GL-parameter}
In mirror chains the magnetic field penetrates the quasi-superconducting region just up to a length $\lambda_m\approx\kappa_G\lambda_L\gtrsim\lambda_L$ with London skin depth $\lambda_L\approx\mathcal{N}^{-1/2}\lambda_i$. Here $\lambda_i=(m_i/m_e)^{1/2}\lambda_e$ is the ion skin depth, $\lambda_e=c/\omega_e$ and $\omega_e$ the electron plasma frequency, and $\kappa_G$ the Ginzburg ratio, defined below. In real mirror bubbles it does, however, deplete the magnetic field only partially, not completely, a point which is fundamental to the above mechanism of phase transition and the generation of chains of mirror bubbles. Maintenance of a magnetic field fraction is crucial because it must maintain and enable the required bounce motion of electrons. The discrete-particle resonance is only temporary and resolves after a while but the large number and distribution of bouncing electrons over the whole volume guarantees for the permanent presence of a locked electron population and the  condensate distributed over the volume of the mirror bubble. The property of a second-kind quasi-superconductor is provided by the Ginzburg-Landau ratio 
\begin{equation}
\kappa_G=\frac{\lambda_L}{\xi}>1
\end{equation}
The plasma perforates into a large number of bubbles (mirror chains) with local diamagnetism caused  by the condensate in each bubble. It does not embrace the whole plasma volume. (Clearly, complete Meissner effects in space, for instance the magnetosheath, are unrealistic as they would deplete the entire plasma volume of magnetic fields on the large scale, which is not observed and thus does not take place.) 
Since $\xi$ is the scale where the electrons feel their mutual attracting potentials close to all the continuously distributed mirror points of the trapped bouncing electrons, it is a natural correlation length of the electrons in the mirror mode plasma. Clearly, the correlation length $\xi\approx1.5\lambda_D<\lambda_m$. In the magnetosheath one then has about $10^2<\kappa_G<10^4$ which suggests strong magnetic perforation as is clearly observed and, in addition large skin depth, reflecting that the bubbles are only partially depleted of the magnetic field. 

In this view mirror mode chains can be considered classical representations of a second-kind superconducting Ginzburg-Landau phase transition from normal to perforated plasma state in high temperature plasma. Their observation in the turbulent magnetosheath behind the bow shock, which is a strong shock, is due to the capacity of the shock to generate conditions in the transition region between the shock and magnetopause which satisfy both the necessary and sufficient conditions for the evolution of the mirror mode. 

 Once mirror chains have evolved and the plasma has become perforated by the quasi-superconducting phase transition described above, the question arises whether the closely spaced mirror bubbles may interact. In the following we focus on this interaction between mirror bubbles and its possible observational signature.

\section{Josephson effect in mirror modes}

The problem of interaction of two superconductors (in our case  two quasi-superconducting partially magnetic field depleted mirror bubbles separated by a non-superconducting magnetized sheet)  is the celebrated Josephson problem \citep{josephson1962,josephson1964}. It makes use of the Landau-Ginzburg mesoscopic theory of superconductivity \citep{ginzburg1950} which is applicable in this case. The order parameter is  the expectation value of the wave function $\psi$ given by $\langle\psi\psi^*\rangle=\mathcal{N}$, which in our case is the above introduced fractional density $\mathcal{N}$  of the bouncing electrons in resonance with the ion-acoustic thermal background fluctuations which form the condensate. The interaction includes of course the boundaries of the two bubbles and hence takes into account the current while the interior  is of little interest in the interaction. It just responses by the exponential partial Meissner screening of the magnetic field $B=B_0\exp(-x/\lambda_m)$ with $\lambda_m$ the penetration depth which for $\mathcal{N}<1$ is larger than the inertial length $\lambda_i=c/\omega_i\approx 43\lambda_e$ and the London length in a proton plasma.  Since mirror bubbles are ion modes even though driven unstable by electrons one here must use the ion inertial length, with $\omega_i$ the ion plasma frequency. Observations suggest that the mirror penetration length is roughly $\lambda_m\approx (10-20)\lambda_L$ which on its own suggests that $\mathcal{N}\lesssim 10^{-2}$.

The wave function of superconduction which in the above spirit we apply to the case of mirror modes obeys the above used first Ginzburg-Landau equation \citep[as was proposed in][]{treumann2018,treumann2019}. The current, being purely electronic, is given by the well known quantum mechanical expression \citep{ginzburg1950,bardeen1957}, the second Ginzburg-Landau equation
\begin{eqnarray}\label{eq-curr}
\mathbf{j}(\mathbf{x},t)&=&-\frac{ie\hbar N_0}{2m_e}\big(\psi^*\nabla\psi-\psi\nabla\psi^*\big)-{|\psi|^2}\frac{N_0}{\kappa_G}\frac{\mathbf{A}}{\mu_0\lambda_e^2}\\
&=&\frac{e\hbar N_0}{m_e}|\psi|^2\Big(\nabla\phi-\frac{e}{\hbar\kappa_G}\mathbf{A}\Big)
\end{eqnarray}
with $\mathbf{A}$ the magnetic vector potential, $\phi$ the phase of the complex wave function, and boundary condition that the normal current must be continuous
\begin{equation}
\mathbf{n}\cdot\Big[\big(\hbar\nabla-\frac{ie\mathbf{A}}{\kappa_G}\big)\psi\Big]_{2-1}=0
\end{equation}
where the brackets mean the difference between the quantities to both sides of the boundary, as indicated by the subscript $2-1$, and $\mathbf{n}$ is the normal to the boundary. Clearly, in a purely classical treatment only the last term in the current expression survives when putting $\hbar=0$. In our semiclassical approach we retain the quantum part of the current, which is the Josephson approximation. Classically the quantum part is neglected, and one has
\begin{equation}
\mathbf{j}(\mathbf{x})=-|\psi|^2\frac{N_0}{\kappa_G}\frac{\mathbf{A}}{\mu_0\lambda_e^2}
\end{equation}
and from Mawell's equations trivially
\begin{equation}
\Box\mathbf{A}=-|\psi|^2\frac{N_0}{\kappa_G}\frac{\mathbf{A}}{\mu_0\lambda_e^2}
\end{equation}
whose solution in one dimension only is clearly $B=B_0\exp(-x/\lambda_m)$, the explicit partial Meissner  skin effect caused by the condensate $|\psi|^2$. (Note again that the current does not explicitly depend on mass, which implies that a hypothetical ion condensate would as well contribute to the phase transition.) Thus the important physics is contained in the generation of the condensate as described in the previous section. If assuming $B/B_0\approx 0.5$ as is typical in the magnetosheath, one has that $\lambda_m\approx 1.5 x$, where $x$ is the measured penetration length. This, in the magnetosheath, is about $x\approx 100$ km. Hence, with $\lambda_i\sim 10$ km one confirms the order of magnitude of $\lambda_m$ as given above.

Now, when  in contrast to the above semi-classical use of the first Ginzburg-Landau equation considering the interaction of the mirror bubbles, the quantum property of the phase has to be retained because it is just the phase which contains the microscopic information. Moreover, space plasmas are ideal conductors and no resistors. Hence the normal current will naturally be different from zero and will reflect the microscopic effect of the interaction. For this reason the quantum part of the current must be retained. We will see that this is important in the case under consideration.

There is, however, a difference in the region between the two bubbles. It is void of any condensate and thus that narrow domain is void of the Meissner effect. The magnetic field and density it contains are spatially constant. Hence the difference between the two regions is just in the quantum mechanical term in the boundary condition and thus cannot be neglected while the conditions in the two adjacent bubbles may be different. Moreover, the tangential currents (which we do not consider here as they contribute to the partial Meissner effect but are not involved into the normal current which must by itself be continuous) flowing in the adjacent bubble boundaries are in opposite directions. This implies that the two bubbles do not merge. They do not attract each other because of the repulsive Lorentz forces such that they remain separated. Nevertheless one may assume that the separation is narrow with non-compensating currents. Since all regions are conducting a normal current will flow. In real superconductors separated by insulators electron tunnelling takes care of normal currents. Here these currents are real. Nevertheless because of the retained quantum mechanical part of the current, Josephson conditions apply to both its sides. These are given as
\begin{eqnarray}
\partial_n\psi_1-\frac{ie}{\hbar\kappa_G}A_n\psi_1&=&b\psi_2\\
\partial_n\psi_2-\frac{ie}{\hbar\kappa_G}A_n\psi_2&=&-b\psi_1
\end{eqnarray}
with $b=$ const some real constant whose value is only of secondary importance here. This means that seen from each bubble's side the effect of the other on the transition is constant and opposite. It is thus assumed that the transition layer between the bubbles is thin enough to consider a constant current when crossing it, thereby for simplicity neglecting any spatial fluctuation or divergence of the current. This may hold as long as the transition distance is short compared with the bubble diameter, a condition satisfied in the magnetosheath, for instance. Moreover, the layer is neither an ideal conductor nor an ideal insulator such that current flow across it is permitted. In the case when it is an ideal conductor it should be narrower than the skin depth outside the mirror bubbles, but even if this does not apply current flow is permitted anyway. 

Inserting these boundary conditions into the current Eq. (\ref{eq-curr}) and cancelling some terms yields for the perpendicular current crossing the thin layer that
\begin{equation}\label{eq-curr-1}
j_n=-\frac{ieb\hbar N_0}{2m_e}\big(\psi_1^*\psi_2-\psi_1\psi_2^*\big)=j_J\sin(\phi_2-\phi_1)
\end{equation}
where $j_J=(eb\hbar/m_e)N_0|\psi|^2$ is the Josephson current, and we for simplicity assumed symmetry $|\psi_1|=|\psi_2|$ up to the phases. (One may note that this current is a pure electron current; any ion-condensate contribution can be neglected because of the inverse proportionality to the mass. Clearly this is an effect of the large electron mobility.) This is most easily seen when replacing the functions $\psi_{1,2}$ by the sum of their real and imaginary parts $\psi=\psi_r+i\psi_i$. The difference in the round brackets is the phase difference between the two wave functions $\psi=|\psi|e^{i\phi}$ in the two bubbles which distinguishes them. This is a retained quantum effect even in the macroscopic case. Its importance comes into account when remembering that the gauge potentials are defined only up to additional functions. The vector potential $\mathbf{A}\to\mathbf{A}+\nabla U$ is defined only up to the gradient of a potential $U$, and consequently the electric field $\mathbf{E}=-\nabla V-\partial_t\mathbf{A}$ up to its time derivative $\partial_tU$. This implies from Eq. (\ref{eq-curr}) that the phase changes as $\phi\to\phi+eU/\hbar\kappa_G$ and the electric potential as $V\to V-\partial_tU$ or after comparison and elimination of $U$
\begin{equation}
\partial_t\phi=eV/\hbar\kappa_G
\end{equation}
showing that the phase is affected by the gauge potential. 

Before continuing, it is most interesting to reflect about what has happened. In principle the electrodynamic equations are gauge invariant which means that the vector and scalar potentials can be changed by adding particular gauge functions while leaving the fields unchanged. This it true also here. However, by applying an external potential to the two mirror bubbles one fixes one particular gauge. This still does not change anything on the fields, it however breaks the gauge symmetry. By providing the mirror modes with a particular electric potential field they shift to a special gauge, and no other gauge can be chosen anymore. In the following we will see which consequences this produces. 

The gauge is in fact a (Weyl)  gauge like in field theory. Time integration, with applied constant external potential $V$, yields the well known form of the Josephson phase
\begin{equation}
\phi_2-\phi_1=(\phi_2-\phi_1)_0-\omega_Jt, \qquad \omega_J= \frac{e}{\hbar\kappa_G}(V_2-V_1)
\end{equation}
which enters into the exponent of the wave function $\psi$. In the presence of a potential difference $V_2-V_1$ the current (\ref{eq-curr-1}) in the junction consisting of the two mirror bubbles with their common boundary of finite thickness will thus oscillate at the Josephson circular frequency $\omega_J$. Correspondingly the normal current becomes
\begin{equation}\label{jn}
j_n=j_J\sin(\Delta\phi_0-\omega_Jt), \qquad \Delta\phi_0=(\phi_2-\phi_1)_0
\end{equation}
This current is a real oscillating classical normal current flowing in the boundary region of the adjacent mirror bubbles. It is a current which varies with time oscillating between the bubbles. (For instance, for $\Delta\phi_0=0$ one has 
\begin{displaymath}
j_n=-j_J\sin{\omega_Jt}
\end{displaymath}
showing that the normal current in the boundary oscillates spatially back and forth between the two bubbles.) Denoting the potential difference as $V_2-V_1=\Delta V$ and introducing the magnetic flux quantum $\Phi_0=\pi\hbar/e$, the observable Josephson frequency becomes
\begin{equation}
\nu_J=\frac{\Delta V}{\Phi_0\kappa_G}, \qquad \Phi_0=2\times 10^{-15}~~\mathrm{Vs}
\end{equation}
This frequency corresponds to an energy $\kappa_G\hbar\omega_J\approx1$ eV multiplied by the applied potential difference $\Delta V$. For $\Delta V\approx 1$ V the oscillation frequency is $\kappa_G\nu_J\approx 5\times10^{14}$ Hz, which is in the near-optical infrared. The applied potential is measured in units of the elementary magnetic flux. It thus reflects the oscillations or transport of elementary flux tubes at high frequency, perfectly suited to measure very small potential differences as used in Josephson SQUIDs.

\section{Radiation}
This result discovered by Josephson \citep{josephson1962} is remarkable as, according to the above discussion, it also occurs under semi-classical conditions if only an electric potential $V(t)$  is applied to the two adjacent mirror bubbles. This is the case in a streaming plasma with plasma flow across the magnetically depleted region or where an externally applied cross potential exists. Examples are the magnetosheath \citep{lucek2005} or other regions like,  for instance, mirror mode chains in the solar wind \citep{winterhalter1994,zhang2008}. Other examples are collisionless shocks \citep{balogh2013} which have comparably narrow transition scales, develop current sheet overshoots between magnetic depletions resembling a similar kind of junctions. Ion-inertial scale plasma turbulence or flow-driven reconnection are further examples.   

The Josephson frequency of oscillation is comparably high. Its large value is due to retaining the quantum effect which implies normalization of the potential difference $\Delta V$ to the small flux quantum $\Phi_0$. Thus the Josephson current oscillates at high frequency. For any classical macroscopic process it averages out as neither the flow nor the density can follow the fluctuation. It is just the current whose real phase oscillates between the two adjacent mirror bubbles. The quantum effect on the macroscopic plasma behaviour of the fields thus disappears. This is, of course, what is expected if considering the flow or evolution of the magnetic field. 

However, there is one effect which is retained even in classical physics. This is radiation which can, in principle, be observed even though its cause is to be found in quantum physics. In this sense the Josephson effect and the frequency resemble the generation of electromagnetic radiation by atomic processes, which are pure quantum effects with macroscopically measurable consequences. In close similarity the Josephson radiation can, in principle, be observed from remote by monitoring its radiation.     

Oscillating currents represent sources of electromagnetic radiation, as prescribed by the wave equation 
\begin{equation}
\Box\,\mathbf{A}_{rad}(\mathbf{x},t)=-\mu_0\,\mathbf{j}(\mathbf{x},t)
\end{equation}
where $\mathbf{A}_{rad}$ is the radiated vector potential, and $\mathbf{j}=j_n\mathbf{n}$, the sinusoidal real current Eq. (\ref{jn}), where it should once more be noted that this is a real classical current. Any natural system which acts, even semi-classically, like a Josephson junction should therefore emit electromagnetic radiation at frequency around $\nu_J=\omega_J/2\pi$. The spectral width of the radiation depends on the spectral width $\Delta\omega$ of the applied time dependent external potential $V(t)$, whose Fourier transform is
\begin{equation}
V(\omega)=\int dt\,V(t)e^{i\omega t}
\end{equation}
whose width can be quite large, compared with the theoretical sharpness of the Josephson frequency, in particular when the flow is highly turbulent. It leads to a time dependent Josephson phase 
\begin{equation}
\phi_J(t)=\Delta\phi_0-\frac{e}{\hbar\kappa_G} \int_0^t dt'\, V(t')
\end{equation}
and, consequently, to a radiation spectrum of some typical width $\Delta\nu$ in frequency. An electric oscillation spectrum $V(\omega)$  yields for the fluctuating part of the phase
\begin{equation}
\phi_J(t)-\Delta\phi_0=-\frac{e}{2\pi i\hbar\kappa_G} \int \frac{d\omega}{\omega}\,V(\omega)\big(e^{-i\omega t}-1\big)
\end{equation}
which, for the realistic case of comparably low frequency oscillations $\omega t\ll 1$ yields that the Josephson frequency becomes
\begin{equation}
\nu_J\approx \frac{1}{4\pi\kappa_G\Phi_0}\int d\omega\,V(\omega)\approx\frac{V(\omega_{max})}{4\pi\kappa_G\Phi_0}\Delta\omega
\end{equation}
The emission spectrum is only as broad as $\Delta\omega$. Since the electric potential arises from external motions in plasma its spectrum is limited from above by the plasma frequency $\Delta\omega\lesssim\omega_e\ll\omega_J$. As the plasma frequency is low, any radiation of Josephson frequency $\nu_J$ will therefore remain to be of very narrow bandwidth $\Delta\nu\ll\nu_J$ and thus, when observed, will form a narrow emission line. Any spectral broadening would then have other reasons having to be retraced to the angle or the tininess of the applied potential. One does of course not expect that this kind of radiation would be intense. Its intensity per unit volume and frequency is well known \citep{jackson1975} to be proportional to the average spectral square of the radiated vector potential, 
\begin{equation}
\frac{dI}{d\Omega d\omega}\propto |A_{rad}(\omega)|^2\propto \mu_0^2|j_J(\omega)|^2=\mu_0^2(eb\hbar/m)^2N_0^2\mathcal{N}^2
\end{equation}
Because of the weakness of the maximum Josephson current $j_J$, any susceptibly high enough radiation intensity requires a  large number of emitters closely distributed in the volume, i.e. a large number of interacting mirror bubbles and hence large volumes, a condition that is probably not realistic in near-Earth space but may sometimes be realized under astrophysical conditions. Hence there should be little doubt about the presence of such a radiation effect, its detectability might however be questionable.

\section{Examples}
\subsection{Streaming mirror mode plasmas}
The case of a streaming plasma is of particular interest. Let the plasma, like that in the magnetosheath, flow at a convection speed $\mathbf{v}$ such that in a frame stationary with respect to the plasma the measured electric field is $\mathbf{E}=-\mathbf{v}\times\mathbf{B}$, where $\mathbf{B}$ is the local magnetic field. Then we have dimensionally
\begin{equation}
\Delta V\approx \frac{v\Phi}{\Delta L}\sin\theta
\end{equation}
where $\Delta L$ is the typical length scale, $\Phi=N_\Phi\Phi_0$ the magnetic flux in the magnetic field, $N_\Phi$ the usually large total number of flux elements, and $\theta$ the angle between the velocity $\mathbf{v}$ and the magnetic field $\mathbf{B}$. The Josephson frequency then becomes
\begin{equation}
\nu_J=\frac{v\Phi}{\Delta L\Phi_0\kappa_G}\sin\theta \approx \frac{vN_\Phi}{\Delta L\kappa_G}\sin\theta~~\mathrm{Hz}
\end{equation}
As for an example, in the magnetosheath we have $v\gtrsim 10^4$ ms$^{-1}$, $B\gtrsim 10$ nT, and $\Delta L\sim 10^4$ m, which gives $\kappa_G\nu_J\gtrsim5\times10^{14}\sin\theta$ Hz just the above estimated frequency. This may become substantially reduced when the angle of plasma flow is close to parallel. 

A texture of mirror bubbles closely spaced to each other  in the magnetosheath should thus glow in the infrared, a frequency which can, without any problem leave the region of its excitation. Mirror mode chains in the solar wind on the other hand are roughly perpendicular to the flow and of generally larger extension. Hence their frequency will be higher closer to the optical range in the very near infrared where they occasionally could be observed. However they seem to occur rather rarely which is in contrast to the region behind shocks like the bow shock. Here they seem to be present almost at any time. 

Similarly one expects that the heliosheath region behind the heliospheric termination shock evolves into magnetic turbulence where the mirror mode will constitute its lowest frequency contribution. The flow speed of the solar wind will become reduced to values similar to the magnetosheath, while the magnetic field drops to $B\approx 0.14$ nT, as expected roughly by two orders of magnitude \citep{burlaga2016,fichtner2020}. The unknown length scale will partially compensate for this drop. One may thus expect that the heliosheath in the region where the mirror mode will become excited behind the termination shock will also glow in the infrared, possibly however at slightly longer wavelengths. In general, any infrared glow around stellar winds might indicate the position of their external boundaries by this kind of Josephson effect which evolves solely in mirror mode turbulences. For very weak magnetic fields or otherwise reduced magnetic fields or slower speeds the frequency of this glow may drop into the microwave domain.

\subsection{The case of reconnection}
In this context it is of particular interest to refer to reconnection as these results are independent of the direction of the magnetic field. Reconnection and mirror modes may be closely related \citep{phan2005,volwerk2003} as one can easily imagine that mirror modes when encountering an antiparallel magnetic field could ignite reconnection. Moreover, they may also evolve in the reconnection process as recent MMS observations \citep{hau2020} of reconnection at the magnetopause and Grad-Shafranov reconstructions suggest.

Consider the case of reconnection when two plasmas of oppositely directed magnetic fields approach each other. Let them be separated by a non-magnetic plasma which by definition is ideally conducting. Then the magnetic fields penetrate it only up to their skin depth $\lambda_e$. When the two plasmas approach each other the nonmagnetic sheet between them will become compressed and ultimately some plasma will be squeezed out into jets escaping from the compressed region in all directions but ultimately preferably parallel/antiparallel to the two external magnetic fields. When the two plasmas are roughly $\sim2\lambda_e$ apart, the separating sheet is a field-free superconductor in whose center two antiparallel exponentially weak fields get into contact and merge. However, the current normal to the boundaries of the sheet will be a Josephson current and oscillate at frequency $\nu_J$ in the potential $\Delta V$ of the approaching plasmas. Hence the reconnection region should radiate at high frequency. If many such regions contact in a large volume the volume may glow in the emitted frequency. Since the plasma is highly diluted, there is no problem for the radiation to escape and become visible from remote if only being composed of many such radiators such that the volume emissivity becomes susceptibly large. A single reconnection region will of course emit very weak radiation only. 

This is probably the case in completely evolved low frequency plasma turbulence. Recently \citep{treumann2015} we suggested that  that the main energy dissipation in fully developed plasma turbulence may be provided at the shortest (electron) scales $\lambda\sim\lambda_e$ by reconnection in the turbulently generated small-scale current vortices into which the streaming turbulent plasmas will necessarily decay on these scales. In general this reconnection in each single small-scale (microscopic) current sheet is weak. However since there are very many such current vortices distributed over the large-scale turbulent volume the integrated dissipation will become substantial such that it under stationary conditions will balance the mechanical energy input at large scales by the large scale flow of the plasma. Since such plasmas are filled by a multitude of small-scale reconnection regions each of them representing a magnetic field-free small region, adjacent reconnection sites represent Josephson junctions and thus should radiate at the local Josephson frequency which when measured provides direct information about the reconnection potential $\Delta V_{rec}$, a quantity which is highly desired to know. 

We note finally that observation radiation from reconnection and/or measuring the Josephson frequency when crossing a reconnection site provides a direct measurement of the reconnection potential which otherwise is nearly impossible to determine for the enormous complexity of the reconnection process and reconnection site. Since the reconnection potential is a most interesting quantity, it would be worth the effort to measure it. The Josephson effect could provide such a possibility by putting a SQUID onto a spacecraft or otherwise trying to measure radiation in the infrared from reconnection sites.

\subsection{Thermal background effects}
In the magnetosheath like in any other high temperature plasma mirror modes are embedded into a relatively intense thermal background of ion-sound fluctuations \citep{gurnett1975,lund1996}. The mean thermal level of these fluctuations \citep{treumann1997}, assuming an isotropic Maxwellian background \citep{krall1973,baumjohann1996}, is 
\begin{equation}
W_{s}\equiv\big\langle{\textstyle\frac{1}{2}}\epsilon_0|\delta E|^2\big\rangle \approx \frac{T_e}{\lambda_D^3}
\end{equation}
where $\lambda_D$ is the Debye length. At temperature $T_e\approx 30$ eV, density $N_0\approx 10^7$ m$^{-3}$ this gives an average electric fluctuation amplitude level of $\langle\delta E\rangle\approx 3\times10^{-5}$ Vm$^{-1}$. Since $\langle\delta E\rangle\approx \langle\Delta V\rangle\Delta L^{-1}$ this yields a Josephson frequency
\begin{equation}
\kappa_G\nu_J\gtrsim 10^{10}\Delta L ~~\mathrm{Hz}
\end{equation}
If $\Delta L\approx 10 \lambda_D\sim 10^2$ m is a typical wavelength of the ion sound noise, then the Josephson frequency is in the range $\sim 10^{3}$ GHz, and the radiation produced is of wavelength $\lambda\sim10^{-4}$ m, in the far infrared or millimeter radio wavelength range. For larger $\Delta L$ it again shifts closer to the infrared.  Considerations of this kind may be of interest in astrophysical objects, in particular in regions of high streaming velocities across collisionless shock waves. Observations in the infrared and short wavelength radio wave spectrum could  provide information about its origins.

On the other hand radiation in some frequency domain may provide information about the potential difference $\Delta V$ of natural systems where either mirror modes evolve or the semi-classical superconducting Meissner effect is generated. For instance radiation at $\hbar\omega_J=100$ eV in the X-ray domain caused by the Josephson effect indicates the presence of potential differences of the order of $10^2$ V. These may not necessarily belong to very strong electric fields, as their strength depends on the scale of the potential differences. 

\subsection{Remarks on shocks}
In collisionless shocks this may indeed be of particular interest. They separate regions of vastly different magnetic field strengths while on the scales of the shock remain collisionless. Moreover, shock fronts exhibit various regions of different properties with spatially highly variable magnetic fields evolving into overshoots and, relative to the shock also ``holes'' in both quasi-perpendicular and quasi-parallel shocks \citep{balogh2013}. Some of these regions may well be considered of similar properties as natural Josephson junctions. Since there the cross shock flow naturally applies a substantial electric potential difference $\Delta V\neq0$, shock transitions of such properties, in particular when relativistic \citep{bykov2011} should become visible as sources of soft X-rays becoming emitters, as is frequently observed in astrophysics. Such X-ray radiation is conventionally attributed to shock acceleration of electrons when interacting with the shock front. However, part of the radiation may also be caused by the Josephson junction effect inside the shock as well as in any texture of mirror modes downstream of the shock. The distinction between the two regions is given by a difference in the emitted spectrum. The much higher velocity difference prevalent to the shock transition than downstream places the former into the X-ray domain, while downstream radiation would be substantially softer reaching into the optical to infrared spectral ranges. Of course, in any case the intensity of the radiation will be low, depending on the number of mirror bubbles, the velocity of the cross flow, and magnetic field. On would expect the highest intensity from strong relativistic shocks or otherwise from a large volume of turbulence.

In the magnetosheath, the region where near Earth one observes mirror modes, the conditions are that the plasma is dilute of the order of $N_0\sim10^6$ m$^{-3}$. The responsible applied electric field is that of the streaming plasma which in mirror modes is quite slow, of the order of $v\approx 10^4$ ms$^{-1}$ in a magnetic field of some $B\approx 10$ nT. This yields an electric field of the order of $E\approx Bv\sin\theta \sim 10^{-4}\sin\theta$ Vm$^{-1}$. This gives an oscillation frequency $\nu_J\sim 10^{10}\Delta L$ Hz, where $\Delta L$ is the width of the sheet crossed by the magnetosheath flow in meters. It probably compensates for the reduction of the frequency, but if the flow is mostly parallel to the magnetic field, then the oscillation frequency may be reduced substantially. A nearly parallel flow may nevertheless bring it down into the microwave domain of GHz. In any case, the mirror mode should become a high frequency radiator. Thus, if a comparably large volume is filled with an ensemble of mirror modes it may manifest itself as a source of incoherent radiation at high frequency if the plasma experiences a cross flow. 

\section{Summary}
Following earlier work on condensate formation in magnetic mirror modes we have provided the conditions for a quasi-superconducting phase transition in high temperature plasma, following the linear mirror instability. In this process bouncing charged particles in discrete particle resonance with the thermal ion acoustic background noise lock to the ion sound wave and temporarily escape from bounce motion while generating a large anisotropy. Through production of a weak attracting electric potential in their wakes they give rise to a correlation length $\xi$. The phase transition is governed by the semi-classical GL theory and results in a second kind quasi-superconducting state exhibiting a partial Meissner effect. Since the Ginzburg ratio $\kappa_G>1$ is large, the phase transition perforates the plasma causing a magnetic texture which consists of chains of mirror bubbles. 

We then investigated the interaction of two closely spaced bubbles finding that it can be described as Josephson junctions producing a classical signature in weak high frequency electromagnetic radiation at frequency depending on the equivalent electric field and direction of the plasma flow. Its frequency is far above the plasma frequency cut-off such that it would be observable from remote. Though weak and if observable it maps the mirror mode region into frequency space. Similar effects are expected in reconnection and shocks and could be of interest in application to astrophysical objects.

\begin{acknowledgement}
This work was part of a Visiting Scientist Programme at the International Space Science Institute Bern. We acknowledge the hospitality of the ISSI directorate and staff. We acknowledge helpful discussions with a number of colleagues, A. Balogh, C.H.K. Chen, R. Nakamura,  Y. Narita, Z. V\"or\"os, and others, some of them being rather critical of any detectable  mesoscale or macroscale quantum effects in high temperature plasmas.  
\end{acknowledgement}



\end{document}